\documentclass{article}
\usepackage{spconf,amsmath,graphicx}
\usepackage{placeins}

\usepackage{booktabs}
\usepackage{ascmac}
\usepackage{amsfonts}

\newcommand{\Z}{\mathbf{Z}}
\newcommand{\X}{\mathbf{X}}
\newcommand{\y}{\mathbf{y}}
\newcommand{\Real}{\mathbb{R}}
\newcommand{\Vocab}{\mathcal{V}}
\newcommand{\diff}{\mathsf{diff}}
\newcommand{\reff}{\mathsf{ref}}
\newcommand{\inp}{\mathsf{input}}

\newcommand{\encoder}{\mathsf{AudioEncoder}}
\newcommand{\decoder}{\mathsf{TextDecoder}}

\title{Audio Difference Learning for Audio Captioning}
\name{Tatsuya Komatsu$^{1,2}$, Yusuke Fujita$^{1}$, Kazuya Takeda$^{2}$, Tomoki Toda$^{2}$}
\address{
$^{1}$LINE Corporation, Japan, 
$^{2}$Nagoya University, Japan}
\begin{document}
%
\maketitle
\begin{abstract}
This study introduces a novel training paradigm, audio difference learning, for improving audio captioning. 
The fundamental concept of the proposed learning method is to create a feature representation space that preserves the relationship between audio, enabling the generation of captions that detail intricate audio information. 
This method employs a reference audio along with the input audio, both of which are transformed into feature representations via a shared encoder. Captions are then generated from these differential features to describe their differences. Furthermore, a unique technique is proposed that involves mixing the input audio with additional audio, and using the additional audio as a reference. This results in the difference between the mixed audio and the reference audio reverting back to the original input audio. This allows the original input's caption to be used as the caption for their difference, eliminating the need for additional annotations for the differences.
In the experiments using the Clotho and ESC50 datasets, the proposed method demonstrated an improvement in the SPIDEr score by 7\% compared to conventional methods.
\end{abstract}
\begin{keywords}
audio captioning, audio difference captioning, audio difference learning
\hspace{-10mm}
\end{keywords}
\section{Introduction}
\label{sec:intro}
Audio captioning, the task of describing the content of an input audio in natural language, has emerged as a vital technology in the field of audio processing~\cite{drossos2017automated,wu2019audio,Mei2021act,mei2022survey}. 
This technology serves a multitude of applications, such as providing accessibility services for the hearing impaired, enabling efficient audio content search, and analyzing the audio environment for surveillance systems. 

Many audio captioning techniques currently adopt an encoder-decoder framework~\cite{sutskever2014sequence}.
This system consists of an audio encoder that captures feature representations for captioning from the audio and a text decoder that generates captions from these representations. 
Techniques such as RNNs (Recurrent Neural Networks)~\cite{drossos2017automated,wu2019audio}, CNNs (Convolutional Neural Networks), and Transformers~\cite{vaswani2017attention} are often utilized in the audio encoder~\cite{Mei2021act},
while the decoder frequently employs Transformers.

One of the primary challenges in audio captioning is the limited availability of paired audio-caption data. 
For instance, the widely-used Clotho dataset~\cite{drossos2020clotho} comprises roughly 5,000 audio clips ranging from 15 to 30 seconds, totaling about 30 hours. 
AudioCaps~\cite{kim2019audiocaps} contains about 140 hours of data from 50,000 clips, each around 10 seconds long. 
When compared to Automatic Speech Recognition (ASR) which handles data spanning hundreds or even thousands of hours, audio captioning is confined to a relatively small dataset.
Therefore, many researchers resort to leveraging pre-trained models like PANNs~\cite{kong2020panns} and BEATs~\cite{chen2022beats}, and employ data augmentation techniques.

Data augmentation for audio captioning, however, is not straightforward.
While audio input alone can be processed by lots of techniques such as speed perturbation and SpecAugment~\cite{park2019specaugment}, augmenting captions is more challenging. 
Several methods use rephrasing captions such as synonym substitution~\cite{chang2023_t6a} and utilizing language models~\cite{Primus2023,wu2023_t6a}. 
Recently, methods inspired by MixGen~\cite{Hao2023} proposed in the vision-language domain have been proposed~\cite{kim2023exploring,chang2023_t6a,Kim2023a}.
These involve mixing two audios and concatenating their respective captions with conjunctions like `and'.
Other methods proposed data augmentation using temporal connectors like `followed by' or `after' to handle temporal dependencies of audio contents~\cite{wu2023audio,xie23d_interspeech}. 
However, these rule-based techniques often yield limited effectiveness, sometimes only enhancing performance on specific metrics. The augmented captions may result in a deviation from the actual captioning, risking performance degradation. 
Thus, an effective learning methodology is imperative.
\begin{figure*}\vspace{-6mm}
    \centering
    \includegraphics[width=0.85\linewidth]{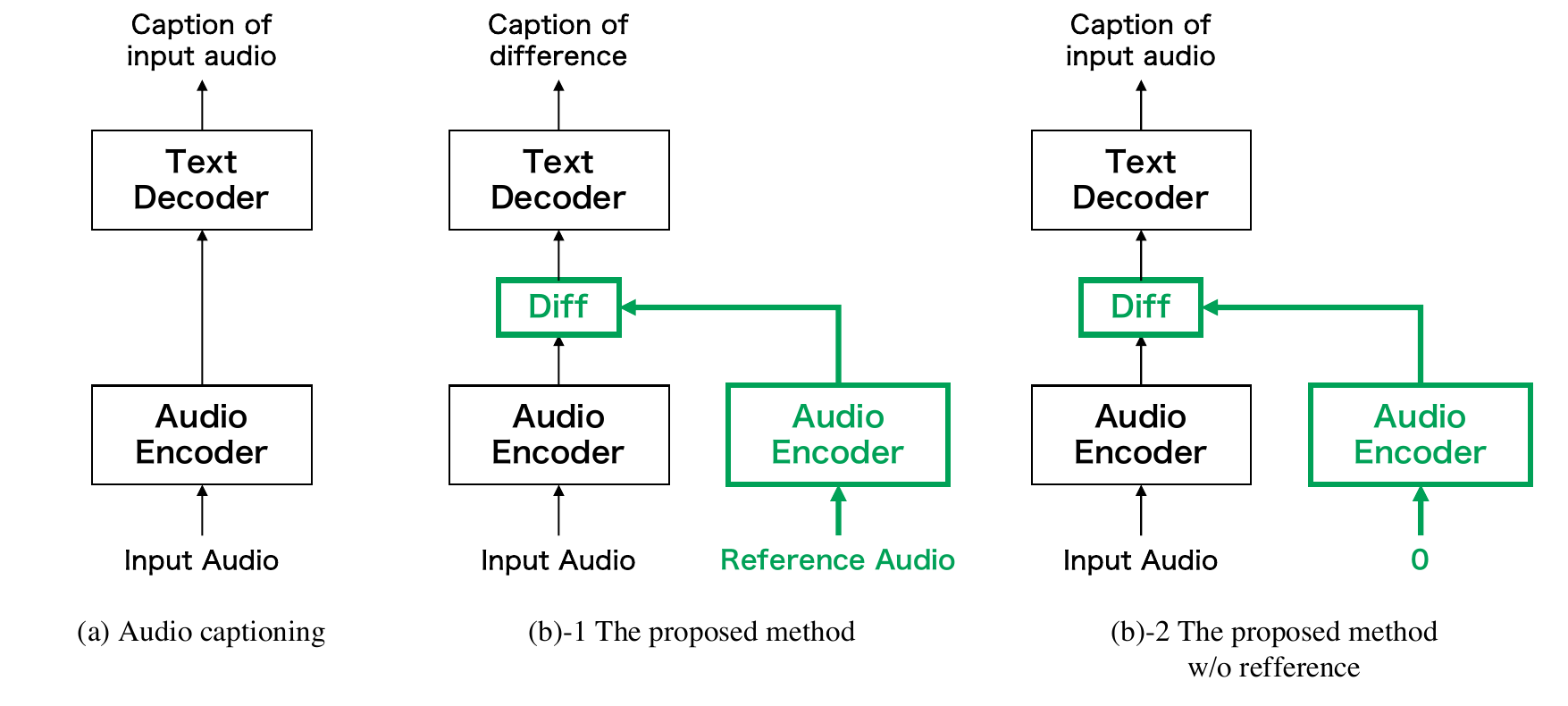}\vspace{-5mm}
    \caption{
(a) A conventional audio captioning system, 
(b)-1 The proposed method generates the difference between the input and reference audio based on the difference of the encoded audio representation.
(b)-2 The proposed method behaves as the general audio captioning system when the reference audio is set to zero.
    }
    \label{fig:threetype}\vspace{-4mm}
\end{figure*}

In this paper, we propose a novel learning approach called `audio difference learning'.
This method introduces a reference audio as an additional input upon training.
We then generate a caption based on the difference between the original input audio and the reference audio.
This difference is computed in the feature space, and captions are generated based on the feature-difference representation.
Moreover, we introduce a learning trick to avoid preparing manual annotations and heuristic-rule-based text processing for audio difference learning.
Specifically, we mix the reference audio with the original input audio, using the resultant mix as a new input audio. 
By calculating the difference between the mixed input audio and the reference audio in the feature space, we can reproduce the feature representation of the original input. 
As a result, the corresponding target caption remains unchanged from the original, enabling training without the need for additional annotations or text processing.
This method also paves the way for new applications where the differences between two audios can be captioned.

\section{Audio Captioning}
Audio captioning is a task that describes the content of input audio through a sequence of text. 
Audio captioning systems generally consist of an audio encoder that transforms input audio into a feature representation
and a text decoder that generates text captions from the obtained feature representation.
A block diagram of this process is shown in Figure~\ref{fig:threetype}-(a).

Let's denote the spectral feature of the input audio as $\X_{\inp}\in\Real^{T\times F}$ and 
the target text sequence as $\y\in\Vocab^{L}$, where $T$ is the length of the audio sequence, $F$ is the number of the frequency bins, and $L$ is the length of the text sequence.
First, the input audio $\X_\inp$ is fed into the audio encoder, transforming it into a feature representation $\Z\in\Real^{T\times{D}}$, where $D$ is the feature dimension.
This transformation can be expressed as:
\begin{align}\label{eq:enc}
    \Z = \encoder(\X_\inp).
\end{align}
The feature representation $\Z$ can be considered as a representation of the semantic content within $\X_\inp$.
By inputting $\Z$ into the decoder, we obtain an estimation of $\y$ as:
\begin{align}
    \hat\y = \decoder(\Z).
\end{align}
Here, $\hat\y\in[0,1]^{|\Vocab|\times L}$ represents the probability distribution of text sequence corresponding to the content of the input audio $\X$.

Cross entropy between the predicted text sequence $\hat\y$ and the target text sequence $\y$ is commonly used as a loss function in the training process:
\begin{align}\label{eq:ce}
    \mathcal{L} = \mathsf{CrossEntropy}(\y, \hat\y)
\end{align}
In addition, it is often the case that performance is enhanced by fine-tuning with reinforcement learning on top of this cross-entropy loss~\cite{Mei2021}.
In this paper, only cross-entropy loss was used.

\section{Proposed method:\\Audio difference learning}
\subsection{Basic Framework}
In this paper, we propose audio difference learning which utilizes an additional reference audio input $\X_\reff\in\Real^{T\times{F}}$.
The proposed method is trained to describe the difference between the input $\X_\inp$ and the reference $\X_\reff$, allowing for the construction of a captioning system that takes into account subtle differences by learning the differences between audios. 

The key aspect of our method is to construct a network that can perform semantic transformations of audio content in the feature space by training a model based on differences in the representations. 
The structure is shown in~Figure~\ref{fig:threetype}-(b).
We start by encoding $\X_\reff$ into a reference feature representation $\Z_\reff$ as in Eq.~\ref{eq:enc}:
\begin{align}
    \Z_\reff = \encoder(\X_\reff).
\end{align}
We then derive a difference representation $\Z_\diff$ by calculating the difference between the input $\Z_\inp$ and the reference $\Z_\reff$:
\begin{align}
  \Z_\diff &= \diff(\Z_\inp, \Z_\reff) 
           = \Z_\inp - \Z_\reff, 
\end{align}
where the difference function $\diff()$ can be any function. In this work, we use simple subtraction.
By training the captioning system based on the difference in feature representations, 
we aim to design the feature space where semantic addition and subtraction of audio can be performed.

The difference representation $\Z_\diff$ is fed into the decoder, 
\begin{align}
    \hat\y_\diff = \decoder(\Z_\diff).
\end{align}
and we obtain $\hat\y_\diff$ which is a caption of the difference between the input and the reference audio. 
The challenge here lies in the difficulty of obtaining ground truth labels of $\y_\diff$ to calculate the cross-entropy loss,
\begin{align}\label{eq:ce_diff}
    \mathcal{L}_\diff = \mathsf{CrossEntropy}(\y_\diff, \hat\y_\diff).
\end{align}
To train using Eq.~\ref{eq:ce_diff}, it is necessary to annotate the text $\y_\diff$ that represents the difference between $\X_\inp$ and $\X_\reff$. 

In this work, we present a learning strategy that circumvents the need for annotation of differences.
We construct a new input $\X_{\inp}^+$ by adding the reference audio $\X_\reff$ to the original input $\X_\inp$:
\begin{align}
    \X_{\inp}^+  &= \X_\inp + \X_\reff
\end{align}
The content of $\X_{\inp}^+$ encompasses information about the reference audio with the original input $\X_\inp$.
The difference $\X_{\inp}^+-\X_\reff$ should only encapsulate information from the original input $\X_\inp$.
Hence, the captioning target of the difference $\X_{\inp}^+-\X_\reff$ should be $\y$, which represents the content of $\X_\inp$ as:
\begin{align}
    \hat\y_\diff^+ &= \decoder(\Z_\diff) \\
    &= \decoder(\Z_{\inp}^+ - \Z_\reff) \\
    &= \decoder(\Z_\inp)    
\end{align}
As a result, we can calculate the cross entropy loss using the caption of the original input, $\y$, as
\begin{align}\label{eq:ce_diff2}
    \mathcal{L}_\diff^+ = \mathsf{CrossEntropy}(\y, \hat\y_\diff^+).
\end{align}
This approach enables training the model based on audio differences without the need for additional annotation costs.

Note that if the reference audio $\X_\reff$ is set to zero, the system behaves identically to a conventional audio captioning system as shown in Figure~\ref{fig:threetype}-(b).
This showcases the versatility of our proposed method, making it applicable to various scenarios and inputs.

\subsection{Design of Training Dataset}
In our experiments, we utilized two datasets: the Clotho~\cite{drossos2020clotho} and ESC-50 datasets~\cite{piczak2015dataset}. The Clotho dataset is commonly used in audio captioning tasks and consists of 4981 audio clips, each 15-30 seconds long. These clips are divided into 2893 for training, 1045 for validation, and 1043 for testing.
The ESC-50 dataset is a collection of 2000 environmental sound clips evenly distributed across 50 different classes. Each class represents a specific sound event, providing a range of reference sounds for our study.

For the generation of $\X_{\inp}^+$, we superimposed audio clips from the Clotho and ESC-50 datasets in time domain at the same power. This process involved adding individual sound events from the ESC-50 dataset to the audio scenes from the Clotho dataset. 

By using individual sound events as references, we aimed to facilitate a more granular learning of the latent space, compared to using a captioning dataset with a mixture of diverse sounds. The combination of these two datasets allowed us to effectively test the performance of our proposed method in discerning and quantifying the differences between various audio inputs.
\vspace{-2mm}
\section{Relation to the Prior Studies}
Several data augmentation methods have been proposed for audio captioning. 
For instance, some approaches~\cite{kim2023exploring,chang2023_t6a,Kim2023a}, inspired by the MixGen~\cite{Hao2023} method in the vision-language domain, concatenate captions for mixed audio.
Another approach utilizes a rule-based caption composition that captures the temporal structure of audio~\cite{wu2023audio,xie23d_interspeech}.  While these methods are effective in capturing the temporal structure, they tend to degrade the performance of standard captioning metrics due to the simplicity of the heuristic rule-based composition.

Another approach utilizes ChatGPT to create a mix of two captions~\cite{wu2023_t6a} or to rephrase existing captions~\cite{Primus2023}. While this method can diversify the caption data, it does not inherently consider the audio content, potentially leading to discrepancies between the audio and the generated captions.

Regarding the audio difference, which is the focus of our proposed method, several methods have been proposed to describe the difference~\cite{Tsubaki2023, Takeuchi2023}.
These methods aim to caption the difference between two audios, and they are not intended to improve the performance of standard audio captioning.
Furthermore, they require special captions that represent the difference between two audios for training.
\begin{table*}[!htb]\vspace{-6mm}
\centering
\caption{Experimental results of the general audio captioning task setting: The proposed method employed the reference audio only during the training phase, and it was not used during the evaluation. These results highlight the impact of our proposed audio difference learning on the general audio captioning.}
\label{tab:results}
\resizebox{1\textwidth}{!}{%
\begin{tabular}{@{}lllllllllll@{}}
\toprule
method                                   & bleu$_1$  $(\uparrow)$ & bleu$_2$ $(\uparrow)$  & bleu$_3$ $(\uparrow)$  & bleu$_4$ $(\uparrow)$  & meteor  $(\uparrow)$ & rouge$_l$ $(\uparrow)$  & cider $(\uparrow)$  & spice $(\uparrow)$  & spider $(\uparrow)$  & spider$_{fl}$ $(\uparrow)$  \\ \midrule
Baseline & 
0.576 & 0.381 & 0.256 & 0.166 & 0.180 & 0.384 & 0.420 & 0.123 & 0.272 & 0.264 \\
AL-MixGen~\cite{kim2023exploring}   & 
0.585 & 0.385 & \textbf{0.260} & \textbf{0.169} & 0.182 & 0.386 & 0.425 & 0.126 & 0.275 & 0.269 \\
Proposed &
\textbf{0.601} & \textbf{0.389} & \textbf{0.260} & 0.166 & \textbf{0.194} & \textbf{0.388} & \textbf{0.454} & \textbf{0.127} & \textbf{0.291} & \textbf{0.289} \\
\bottomrule
\end{tabular}%
}
\vspace{-0mm}
\end{table*}
In contrast, our proposed method, audio difference learning, is designed to augment the data by learning the differences between audios.
The proposed method not only diversifies the data but also eliminates the need for additional human annotations, offering a cost-effective and scalable solution for data augmentation in audio captioning.

\section{Experiments}
\subsection{Experimental settings}
We employed the baseline system\footnote{https://github.com/felixgontier/dcase-2023-baseline} of the DCASE2023 task6 for the captioning model. The hyperparameter settings were kept the same with the baseline.
The input audio feature is a 64-dimensional mel-spectrogram with a sampling rate of 44.1 kHz, a window length is 40 ms, and a hop size is 20 ms.
The audio encoder uses a pre-trained CNN layer consisting of 12 layers, followed by an adapter layer consisting of linear layers. 
The dimension of the feature representation is 768. The text decoder uses BART~\cite{lewis2020bart}.
The training was conducted for 40 epochs, with a batch size of 32.

In the experiment, a comparison was made between the baseline trained with standard cross-entropy and the proposed audio difference learning.
We also conducted a comparison with Al-MixGen, also known as PairMix~\cite{kim2023exploring}, a mix-up-like augmentation that involves mixing two audio files and creating a target mixed caption by concatenating the individual captions.

The performance of captioning is evaluated using using coco caption toolkit\footnote{https://github.com/tylin/coco-caption} which generates conventional metrics~\cite{mei2022survey} like BLEU$_n$, METEOR, and ROUGE$_L$, which evaluate n-gram precision and word-to-word matching, among other factors. Metrics such as CIDEr, SPICE, and SPIDEr are also employed, focusing on aspects such as term frequency-inverse document frequency and scene graph captions.
\vspace{-2mm}
\subsection{Results}
The experimental results are shown in Table 1. 
The proposed method was found to improve performance on nearly all measures compared to the baseline and mix-up. In the SPIDEr metric, an improvement of 7\% was achieved.
The proposed audio difference learning was confirmed to be effectively learned.

Table~\ref{tab:example} presents a comparison of the captions generated by our proposed method and the baseline with four different way: 
(1) Original input from the Clotho dataset ($\inp$),
(2) Mixed sound of $\inp$ with sounds from the ESC-50 dataset ($\inp_+$), 
(3) Captions generated by calculating the difference representation between the mixed audio and ESC-50 sounds (this should generate the same caption as (1)). 
(4) Captions generated by taking the difference representation between the mixed audio and the original input, which is the inverse of (3), leaving the ESC-50 as the residual to be captioned.

Both the proposed method and the baseline perform well in captioning the $\mathsf{input}$. 
However, the baseline struggles with the $\mathsf{input_+}$ where an event is superimposed. 
The proposed audio difference learning method successfully expresses individual contents of audio in the representation space.

In the examples where captions are generated based on the difference representation, it can be seen that the proposed method is able to caption only the semantic difference from the original audio in the input sound as highlighted with bold.
The baseline, on the other hand, is unable to handle the difference in the feature space effectively, resulting in captions that are a mixture of contents in superimposed sounds, particularly for $\mathsf{input_+}-\mathsf{input}=\mathsf{input_{esc}}$.

The proposed method not only improves performance but also suggests the potential for additional new applications, such as captioning differences between audio. 
\begin{table}[h]
\centering
\caption{Examples of difference captioning results generated using difference-representation between two audio. The proposed method can handle mixed sounds and differences.}
\label{tab:example}
\resizebox{\columnwidth}{!}{%
\begin{tabular}{@{}ll@{}}
\toprule
\toprule
\multicolumn{2}{l}{{\bf{(1) Input Audio}}: \it{kids are playing as one child shrieks while birds are chirping}}                 \\
~~~\bf{Baseline}         & birds are chirping and children are talking in the background                           \\
~~~\bf{Proposed}         & birds are chirping and children are talking and playing in the background               \\ \midrule
\multicolumn{2}{l}{{\bf{(2) Caption for mixed audio}}: $\inp_+ = \mathsf{\inp} +  \mathsf{Car~Horn~sound}$ }                 \\
~~~\bf{Baseline}         & birds are chirping and children are talking in the background                           \\
~~~\bf{Proposed}         & birds are chirping and children are talking in the background {\bf{as a car drives by}}       \\ \midrule
\multicolumn{2}{l}{{\bf{(3) Caption with difference representation}}: $\mathsf{\inp_+} -  \mathsf{Car~Horn~sound} \Rightarrow \mathsf{(1)}$}                 \\
~~~\bf{Baseline}         & birds are chirping and children are talking to each other                               \\
~~~\bf{Propose}         & birds are chirping and children are talking in the background                           \\ \midrule
\multicolumn{2}{l}{{\bf{(4) Caption with difference representation}}: $\mathsf{\inp_+} -  \mathsf{\inp} \Rightarrow \mathsf{Car~Horn~sound}$}                 \\
~~~\bf{Baseline}         & a person is using a hard object to make a few seconds                                   \\
~~~\bf{Proposed}         & {\bf{an engine is whirring}} and then it gets louder and louder                                \\ \bottomrule
\end{tabular}%
}
\resizebox{\columnwidth}{!}{%
\begin{tabular}{@{}ll@{}}
\toprule
\multicolumn{2}{l}{{\bf{(1) Input Audio}}: \it{a distorted drum or similar instrument is played}}                  \\
~~~\bf{Baseline}         & a synthesizer is playing a musical instrument                           \\
~~~\bf{Proposed}         & a synthesizer is playing a synthesizer with a musical instrument               \\ \midrule
\multicolumn{2}{l}{{\bf{(2) Caption for mixed audio}}: $\inp_+ = \mathsf{\inp} +  \mathsf{Laughing~sound}$ }                 \\
~~~\bf{Baseline}         & a person is playing a synthesizer with a musical instrument in the background                           \\
~~~\bf{Proposed}         & a person is playing a synthesizer {\bf{with a man talks in the background}}        \\ \midrule
\multicolumn{2}{l}{{\bf{(3) Caption with difference representation}}: $\mathsf{\inp_+} -  \mathsf{Laughing~sound} \Rightarrow \mathsf{(1)}$}                 \\
~~~\bf{Baseline}         & a synthesizer is playing a musical instrument                               \\
~~~\bf{Propose}         & a synthesizer is playing a musical instrument                           \\ \midrule
\multicolumn{2}{l}{{\bf{(4) Caption with difference representation}}: $\mathsf{\inp_+} -  \mathsf{\inp} \Rightarrow \mathsf{Laughing~sound}$}                 \\
~~~\bf{Baseline}         & a person is playing a synthesizer while another person is speaking in the background                                   \\
~~~\bf{Proposed}         & a person is speaking and then {\bf{a child laughs}}                               \\ \bottomrule \bottomrule
\end{tabular}%
}\vspace{-3mm}
\end{table}
\vspace{-3mm}
\section{Conclusion}
In this paper, we proposed a novel learning method for audio captioning, called Audio Difference Learning. 
Our proposed method trains the model to generate captions from the feature representations of the differences between audios, thereby constructing a space that can represent these differences.
Furthermore, by designing the input and reference audio such that the difference representation can reproduce the original data, it is possible to carry out learning without human-annotation of the differences. 
Experiments demonstrated that our proposed method exhibited superior captioning performance.
Furthermore, it suggested the potential for a new application that generates caption of differences.

\vspace{3mm}
\noindent\textbf{Aknowledgement}
This work was partly supported by a project, JPNP20006, commissioned by NEDO and JSPS KAKENHI Grant Number 21H04892, Japan.

\vfill\pagebreak
\ninept
\bibliographystyle{IEEEbib}
\bibliography{strings,refs}

\end{document}